\documentclass[showpacs,aps]{revtex4}

\usepackage{graphicx}
\usepackage{epstopdf}
\begin{document}

\title{Implications of B meson decay data - a graphical approach}

\author{Timothy Carruthers}\email{timothyc@physics.unimelb.edu.au}
\author{Bruce H J McKellar}\email{mckellar@physics.unimelb.edu.au}
\affiliation{School of Physics, Research Centre for High Energy Physics,The University of Melbourne, Victoria 3010, Australia}

\date{13 December 2004}
\begin{abstract}
The considerable influx of new data on B decays offers opportunities to determine some of the smaller contributions to the decay process, such as suppressed diagrams, and to test the relations between decay amplitudes implied by (broken) SU(3) symmetry. The graphical decomposition of decay amplitudes, corrected here for the contribution of electroweak penguins following Gronau, Rosner et al, implies bounds not only on the size of diagram contributions but also on the relevant complex phases due to Cabbibo-Kobayashi-Maskawa matrix factors.\\
The available data on B decays to two charmless pseudoscalar mesons are analysed to study some of these decay amplitude relations, and allowed values of $\gamma=Arg(V_{ub}^*)$ which are consistent with the current favoured  values is obtained.  Our analysis is based on a graphical approach, which offers some insight that is difficult to obtain from $\chi^2$ minimisation, and our results are consistent with recent observations on the difficulty of obtaining simultaneous fits to $K\pi$ and $\pi\pi$ decay channels.  We have also analyzed the $K\eta$ and $K\eta'$ channels, and find similar difficulties.
\end{abstract}

\pacs{13.25.Hw}
\maketitle

Decays of light $B$ mesons to a pair of charmless pseudoscalars are all at the branching ratio level of $10^{-5}$ and less, so that in many cases significant data has only been available recently (since ICHEP 2004) and is summarised by the Heavy Flavour Averaging Group~\cite{hfag}. It is expected that improving data over the next few years will enable this general approach to be further refined.

In our analysis, we follow the convention of meson wavefunctions as used by \cite{9404283}, \cite{9509325} and others.  For convenience these wavefunctions are listed here:

\begin{itemize}
\item
Beautiful mesons:
\[ B^0=\bar{b}d \; , \quad B^+=\bar{b}d \; , \quad B_s=\bar{b}s \]
\[ \overline{B^0}=b\bar{d} \; , \quad B^-=-b\bar{d} \; , \quad \overline{B}_s=b\bar{s} \]
\item
Pseudoscalar meson octet:
\[ K^0=\bar{s}d \; , \quad K+=\bar{s}u \]
\[ \pi^-=-d\bar{u} \; , \quad \pi^0=(d\bar{d}-u\bar{u})/\sqrt{2} \; , \quad \pi^+=u\bar{d} \]
\[ K^-=-s\bar{u} \; , \quad \overline{K^0}=s\bar{d} \]
\[ \eta_8=(2\bar{s}s-\bar{u}u-\bar{d}d)/\sqrt{6} \]
\item
Pseudoscalar meson singlet:
\[ \eta_1=(\bar{u}u+\bar{d}d+\bar{s}s)/\sqrt{3} \]
\end{itemize}

The physical $\eta $  and $\eta'$  states are an octet-singlet mixture, with a mixing angle we take to be $\sin^{-1}[\frac{1}{3}]\simeq 19.5^\circ$, corresponding to:

\[ \eta=(\bar{s}s-\bar{u}u-\bar{d}d)/\sqrt{3} \]
\[ \eta'=(2\bar{s}s+\bar{u}u+\bar{d}d)/\sqrt{6} \]

The Hamiltonian governing the weak decays of B mesons has the following structure\cite{9810482}.

\begin{equation}
\label{eq:lowordHam}
\mathcal{H}=\frac{G_F}{\sqrt{2}}\sum_{q=d,s}\left(\sum_{q'=u,c}\lambda_{q'}^{(q)}[c_1(\bar{b}q')_{V-A}(\bar{q'}q)_{V-A}+c_2(\bar{b}q)_{V-A}(\bar{q'}q')_{V-A}]-\lambda_t^{(q)}\sum_{i=3}^{10}c_iQ_i^{(q)}\right)
\end{equation}

where $\lambda_{q'}^{(q)}=V_{q'b}^*V_{q'q}$ are products of CKM matrix elements. The $c_i$ are Wilson coefficients, for which we use the leading log values at $m_b$, as calculated by \cite{9512380}:

\[ c_1=1.144 \; , \quad c_2=-0.308 \]
\begin{equation}
\label{wilsons}
c_7=0.045\alpha  \; , \quad c_8=-0.048\alpha  \; , \quad c_9=-1.280\alpha  \; , \quad c_{10}=0.328\alpha
\end{equation}

Rather than write decay amplitudes explicitly in terms of reduced matrix elements (based on the SU(3) transformation properties of the various operators), it is equivalent to construct decay amplitudes based on the possible graphical contributions to decays. This has been discussed at length elsewhere \cite{9404283} \cite{9509325} \cite{9810482} ---  to summarise, there are 4 important diagrams contributing to amplitudes at lowest order:

\begin{enumerate}
\item
A colour-favoured tree diagram $T$ with a complex phase of $\gamma$ (ie, $T=|T|e^{i\gamma}$).
\item
A colour-suppressed tree diagram $C$ with a complex phase of $\gamma$. The proportionality of $C$ to $T$ diagrams has been traditionally assumed to be primarily due to the QCD effects involved, which can be in principle be calculated using factorisation methods. On this basis $|C/T|$ is thought to be between 0.08 and 0.37 \cite{0306021}. However, this is complicated by the fact that `tree-level' operators in the Hamiltonian contain not only pure $T$ or $C$ diagrams but also penguin and exchange elements \cite{9810482}. It is conjectured that interference of these elements is responsible for a situation of $|C/T|\sim 1$ found according to \cite{0312259} \cite{0402112} and also below. 
\item
A gluonic penguin diagram $P$. In the case of a $|\triangle S|=1$ transition ($P'$), the weak phase of this diagram is $\pi=\mbox{Arg}(V_{ts}V_{tb})^*$ \cite{9509325}, while for $|\triangle S|=0$ transitions the equivalent weak phase is $-\beta=Arg(V_{td}V_{tb}^*)$.
\item
Additionally, to account the presence of SU(3) singlet mesons it is necessary to take account of a gluonic `singlet penguin'  diagram $S$ (which is expected to have the same phase as a non-singlet penguin). Regretfully the resemblance of this diagram to a penguin wearing a singlet is minimal. 
\end{enumerate}

All of these amplitudes acquire strong interaction phases, and indeed these phases are necessary for the existence of some recently obseerved CP violating asymmetries.  The phase required may be small --- He and McKellar~\cite{He:2004ck} found a strong phase of about $6^{\circ}$ for $C'$ and $T'$ accounts for the observed CP violation in $B \to K^-\pi^+$ --- and we will neglect these strong interaction phases in this analysis.  However, we note that some other fits to the new data have found larger strong interaction phases~\cite{Buras:2004dc, Baek:2004rp}

Electroweak penguin (EWP) contributions are also present and significant \cite{9504327}, but the form of the Hamiltonian allows them to be neatly accounted for. By describing Hamiltonian operators in terms of SU(3) transformation properties and Wilson coefficients,  it can be shown such that, to a close approximation, the electroweak penguins are proportional to tree and colour suppressed tree diagrams~ \cite{9810482}.  Explicitly,  where $P_{EW}$ and $P_{EW}^C$ are colour-favoured and colour-suppressed EWP respectively:
\[ P_{EW}=\frac{-3}{2}\lambda\kappa C \; , \quad P_{EW}^C=\frac{-3}{2}\lambda\kappa T \] 
where $\lambda$ is the product of the relevant EWP CKM factors (eg, $\lambda_t^{(s)}$ for $|\triangle S|=1$) and $\kappa$ the ratio of relevant Wilson coefficients: 

\begin{equation}
\label{kappa}
\frac{c_9+c_{10}}{c_1+c_2}=\frac{c_9-c_{10}}{c_1-c_2}=\kappa\simeq-1.123\alpha
\end{equation}

 In addition, it can be shown that, whenever the diagrams $T$, $C$, $P$ and $S$ appear in a decay amplitude, the EWP diagrams appear in a fixed combination with them \cite{9509287}. Following this insight we ultimately find the following correspondences (omitting CKM factors):

\begin{equation}
\label{ewpWithOthers}
T \rightarrow T+\frac{-3}{2}\kappa C \; , \quad C \rightarrow C+\frac{-3}{2}\kappa T \; , \quad P \rightarrow P+\frac{1}{2}\kappa C \; , \quad S \rightarrow S+\frac{1}{2}\kappa T
\end{equation}

The above situation is also equivalent to the results obtained when applying colour-favoured and -suppressed electroweak penguins to decay amplitudes based purely on whether they are graphically allowed or not, and assuming a coupling of the photon or $Z^0$ of the EWP roughly proportional to the charge of the quark it couples to. This phenomenological approach to accounting for EWPs was detailed in \cite{9504327}.

Aside from these four primary graphical components of a decay amplitude, there are three other low-order diagrams (Exchange, Annihilation and Penguin Annihilation), but these are all suppressed by physics and their magnitudes are estimated to be well within the error margins of the current branching ratio measurements. The analysis appears to support this approximation, although the analysis of He and McKellar~\cite{He:2004ck} found that a  large annihilation diagram improved the global fit to the $K\pi$ and $\pi\pi$ data.  With improving data in the future one can hope that it will soon be possible (and necessary) to investigate these suppressed diagram contributions in depth.

\begin{figure}[ht]
\begin{center}
\includegraphics{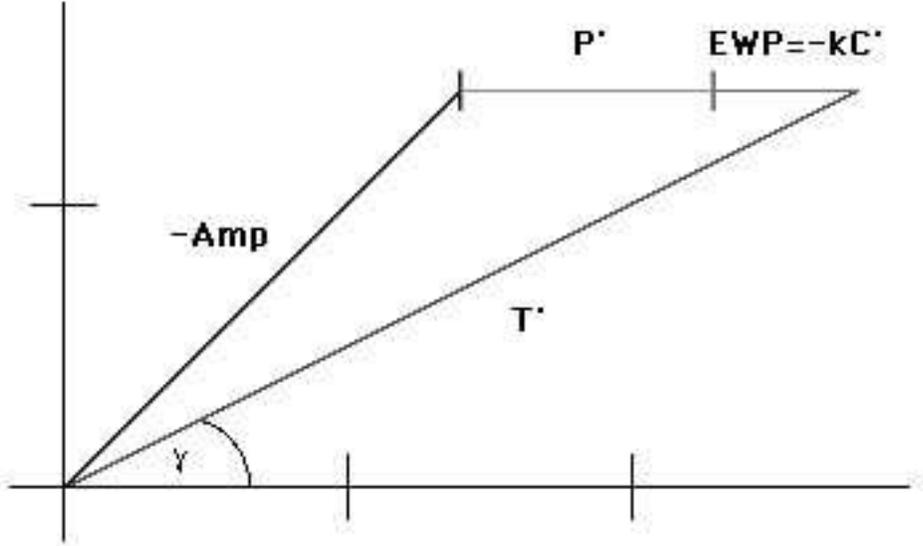}
\end{center}
\caption{\label{k+pi-diag} $B^0\rightarrow K^+\pi^-$ Amplitude ($Amp$) Composition}
\end{figure}

The following table (table \ref{ampstable} below) details the graphical decomposition of $B\rightarrow PP$ decay amplitudes with well-measured branching ratios.
 We follow the convention of \cite{9404283}, where $|\triangle S|=1$ amplitudes are primed, and $|\triangle S|=0$ amplitudes are unprimed.  The data are the averages prepared by the Heavy Flavour Averaging group~\cite{hfag}, and include new data presented at ICHEP 04.\\
\begin{center}
\begin{table}[h]
\begin{tabular}{|l|l|l|c|c|}\hline
Mode&Graphical Amplitudes&EWP Amps$/\kappa$&$BR (\times10^{-6}$)&$|A_{exp}|$ (eV)\\\hline\hline
$B^0\rightarrow\pi^0\pi^0$&$\frac{1}{\sqrt{2}}(-C+P)$&$\frac{1}{\sqrt{2}}(\frac{3}{2}T+\frac{1}{2}C)$&$1.51\pm0.28$ &$13.10\pm1.28$\\
$B^0\rightarrow\pi^+\pi^-$&$-T-P$&$C$&$4.55\pm0.44$&$22.75\pm1.21$\\
$B^+\rightarrow\pi^+\pi^0$&$\frac{-1}{\sqrt{2}}(T+C)$&$\frac{1}{\sqrt{2}}\frac{3}{2}(T+C)$&$5.50\pm0.60$&$23.99\pm1.44$\\
$B^+\rightarrow\pi^+\eta$&$\frac{-1}{\sqrt{3}}(T+C+2P+S)$&$\frac{1}{\sqrt{3}}(T+C)$&$4.80\pm0.60$&$22.53\pm1.53$\\
$B^+\rightarrow\pi^+\eta'$&$\frac{1}{\sqrt{6}}(T+C+2P+4S)$&$\frac{1}{\sqrt{6}}\frac{1}{2}(T-C)$&$4.2\pm 1.1$&$21.31\pm2.91$\\
\hline\hline
$B^0\rightarrow K^0\pi^0$&$\frac{1}{\sqrt{2}}(-C'+P')$&$\frac{1}{\sqrt{2}}(\frac{3}{2}T'+\frac{1}{2}C')$&$11.5\pm1.76$&$36.33\pm 1.76$\\
$B^0\rightarrow K^+\pi^-$&$-T'-P'$&$C'$&$18.16\pm 0.79$&$45.64\pm1.22$\\
$B^0\rightarrow K^0\eta$&$\frac{-1}{\sqrt{3}}(C'+S')$&$\frac{1}{\sqrt{3}}T'$&$2.50 \pm 0.80$&$17.02\pm2.81$\\
$B^0\rightarrow K^0\eta'$&$\frac{1}{\sqrt{6}}(C'+3P'+4S')$&$\frac{1}{\sqrt{6}}(\frac{1}{2}T'+\frac{3}{2}C')$&$65.2\pm6.0$&$87.95\pm4.47$\\
$B^+\rightarrow K^0\pi^+$&$P'$&$\frac{1}{2}C'$&$24.1\pm1.3$&$50.43\pm1.63$\\
$B^+\rightarrow K^+\pi^0$&$\frac{-1}{\sqrt{2}}(T'+C'+P')$&$\frac{1}{\sqrt{2}}(\frac{3}{2}T'+C')$&$12.1\pm 0.8 $&$35.73\pm1.37$\\
$B^+\rightarrow K^+\eta$&$\frac{-1}{\sqrt{3}}(T'+C'+S')$&$\frac{1}{\sqrt{3}}(T'+\frac{3}{2}C')$&$2.60\pm0.50$&$16.65\pm 1.69$\\
$B^+\rightarrow K^+\eta'$&$\frac{1}{\sqrt{6}}(T'+C'+3P'+4S')$&$\frac{1}{\sqrt{6}}\frac{1}{2}T'$&$77.6\pm 4.59$&$91.97\pm3.22$\\
\hline
\end{tabular}
\caption{Amplitude Composition \& Experimental Data}
\label{ampstable}
\end{table}
\end{center}
One can represent decay amplitudes as triangles on a complex plane, formed by the vector addition of their graphical components ($T'$, $C'$ etc), one can obtain relations between the amplitudes (calculated from branching ratios) and their graphical components. As an example the graphical relationship for $B^0\rightarrow K^+\pi^-$ is shown in figure~\ref{k+pi-diag}. Note that this is only very approximately to scale, as the size of partial contributions is not yet determined; also note that for ease of viewing and construction the total amplitude has been rotated by $\pi$.\\

Graphical components (ie amplitude partial contributions) are expected to be the same for all amplitudes in the limit of the SU(3) symmetry of the final-state particles; hence we can use these relations to attempt to bound the major unknown involved, the CKM angle $\gamma$. This is the angle of tree-type diagrams, while in the $|\triangle S|=1$ case penguins and EWP contributions have the same weak phase $\pi$ \cite{9509325} (excluding, as throughout, the possibility of significant strong phase differences, etc). Note that the signs involved in equations \ref{ewpWithOthers} therefore imply that EWP `from' penguins and singlets interfere destructively with them, while EWP `from' tree-type diagrams interfere constructively with any penguins and singlets. 

The most profitable combination of data to consider is the $B\rightarrow K\pi$ decays. Using $B^+\rightarrow K^0\pi^+$ to eliminate $P'$ and its EWP cohort, the component relations remaining involve only the unknowns $T'$, $C'$ and $\gamma$. Furthermore, $T'$ and $C'$ can be further constrained with the amplitude $B^+\rightarrow \pi^+\pi^0$, the composition of which is dependent only on $|T+C|$, $\gamma$ and the CKM angle $\beta$ (measured with good precision at $23.3^\circ\pm1.5^\circ$ \cite{ckmfitterwebsite}).

It is to be expected that the ratio $\frac{|T'+C'|}{|T+C|}$ is directly the basic CKM factor proportion $|V_{us}|/|V_{ud}|$ multiplied by the elementary SU(3) breaking parameter $f_K/f_{\pi}$ (which roughly accounts for the changes due to the $W$ coupling to a kaon rather than pion in the final state). On this assumption we have another piece of data to constrain $\gamma$ in terms of $T'$ and $C'$. The SU(3) breaking ratio of $P'/P$ is not sufficiently certain for it to be useful in this context. For a more thorough investigation of total SU(3) breaking effects in this regime, a variety of new amplitudes is necessary: the interested reader is directed to \cite{9504326}.

Using the graphical representation, we can derive quadratic expressions for $T'$ and $C'$ in terms of $\gamma$, assuming that $T'$ and $C'$ interfere constructively as is expected (and appears to follow from a naive reading of the available data). Our notation is that $[K\pi]=|\mbox{Amplitude}(B\rightarrow K\pi)|$. The graphical amplitudes are defined to include their CKM factors, hence the factor $|\frac{\lambda_t^{(s)}}{\lambda_u^{(s)}}|$, which is necessary to maintain proportion between bare $T'$, $C'$ amplitudes and their proportional EWP contributions. $T'$, $C'$ and $P'$ are assumed to be positive up to their phases following \cite{9404283}, \cite{9409244}, \cite{0306021} etc.\\

\begin{equation}
[K^0\pi^+]=P'+|\frac{\lambda_t^{(s)}}{\lambda_u^{(s)}}|\kappa\frac{1}{2}C'
\label{eq:k0pi+eqn}
\end{equation}

\begin{equation}
[K^+\pi^-]^2=T'^2+(P'-|\frac{\lambda_t^{(s)}}{\lambda_u^{(s)}}|\kappa C')^2-2(P'-|\frac{\lambda_t^{(s)}}{\lambda_u^{(s)}}|\kappa C')T'\cos [\gamma]
\label{eq:k+pi-eqn}
\end{equation}

\begin{eqnarray}
2\times[K^0\pi^0]^2&=&C'^2+\left(P'+|\frac{\lambda_t^{(s)}}{\lambda_u^{(s)}}|\kappa (\frac{3}{2}T'+\frac{1}{2}C')\right)^2 \nonumber\\
&&+2\left(P'+|\frac{\lambda_t^{(s)}}{\lambda_u^{(s)}}|\kappa (\frac{3}{2}T'+\frac{1}{2}C')\right)C'\cos [\gamma]
\label{eq:k0pi0eqn}
\end{eqnarray}

\begin{eqnarray}
2\times[K^+\pi^0]^2&=&(T'+C')^2+\left(P'-|\frac{\lambda_t^{(s)}}{\lambda_u^{(s)}}|\kappa (\frac{3}{2}T'+C')\right)^2 \nonumber\\
&&-2\left(P'-|\frac{\lambda_t^{(s)}}{\lambda_u^{(s)}}|\kappa (\frac{3}{2}T'+C')\right)(T'+C')\cos [\gamma]
\label{eq:k+pi0eqn}
\end{eqnarray}

\begin{eqnarray}
2\times[\pi^+\pi^0]^2&=&(T+C)^2+\left||\frac{\lambda_t^{(d)}}{\lambda_u^{(d)}}|\kappa\frac{3}{2}(T+C) \right|^2 \nonumber\\
&&-2(T+C)\left||\frac{\lambda_t^{(d)}}{\lambda_u^{(d)}}|\kappa\frac{3}{2}(T+C) \right|\cos [\pi-\gamma-\beta]\nonumber\\ \nonumber \\
\rightarrow\frac{|T'+C'|}{|V_{us}|f_K/|V_{ud}|f_\pi}&=&[\pi^+\pi^0]\sqrt{\frac{2}{1+\left||\frac{\lambda_t^{(d)}}{\lambda_u^{(d)}}|\kappa\frac{3}{2}\right|^2+2|\frac{\lambda_t^{(d)}}{\lambda_u^{(d)}}|\kappa\frac{3}{2}\cos[\gamma+\beta]}}
\label{eq:pi+pi0eqn}
\end{eqnarray}

The resultant set of equations above (\ref{eq:k+pi-eqn} to \ref{eq:pi+pi0eqn}), derived from graphical constructions of the relevant decay amplitudes, are not easy to analytically solve simultaneously, although of course a solution can be attempted using Mathematica. However, they can be effectively, and much more transparently, analysed in terms of $T'$ and $C'$ for different (fixed) values of $\gamma$ via simple two-dimensional graphing.

The current best value for $\gamma$ is $(60\pm 14)^\circ$  \cite{PDG}. Graphing the above set of quadratic relations simultaneously (eliminating $P'$ via relation \ref{eq:k0pi+eqn}) for values of $\gamma$ between $38^\circ$ and $80^\circ$, in terms of $T'$ (x axis) and $C'$ (y axis), produced graphs such as that of figure~\ref{58pic}  presented below. As a result of this, it was found that consistent solutions of $T'$ and $C'$ existed only for values of $\gamma$ lower than $58.0^\circ$, using one-standard-deviation ranges for parameters. This is consistent with the currently favoured range, but prefers values less than the current central value.  It provides an important additional constraint on $\gamma$. Additionally, this overlap only allowed for values of $|C'/T'|$ centred around 0.31 (the value necessary at $\gamma=58.0^\circ$), widening steadily to between 0.24 and 0.7 when $\gamma$ reached its allowed lower bound of $38^\circ$. This matches the traditional $|C'/T'|$ bound of between 0.08 and 0.37 due to QCD factorisation (following \cite{0306021}), although it is not entirely inconsistent with the recent work of Buras, Fleischer et al \cite{0312259} \cite{0402112}, and of He and McKellar~\cite{He:2004ck} which suggest a value of $|C'/T'|$ closer to unity.

The graph of the above relations for $\gamma=58.0^\circ$ is shown below, at a scale allowing clear view of all relations (figure \ref{58pic}). Axis units are eV. Note that there are four lines for each of the primary three relations, although not all are visible in this sector of the graph: these are the extent of the $1\sigma$ bounds for the two amplitudes involved in each (the amplitude itself, and the penguin amplitude elimination from relation \ref{eq:k0pi+eqn}). The $[\pi^+\pi^0]$ bound on $|T'+C'|$ is distinguished as a thick band, and the line of $|C'/T'|=0.31$ is also shown. With a value of $\gamma>58.0^\circ$, the $[K^0\pi^0]$ and $[K^+\pi^-]$ no longer intersect within the $[\pi^+\pi^0]$ region. For $\gamma<58.0^\circ$, zones of overlap continue to exist until beyond $38^\circ$, and the necessary value of $|C'/T'|$ widens as $\gamma$ decreases.

\begin{figure}[ht]
\begin{center}
\includegraphics[width=\textwidth]{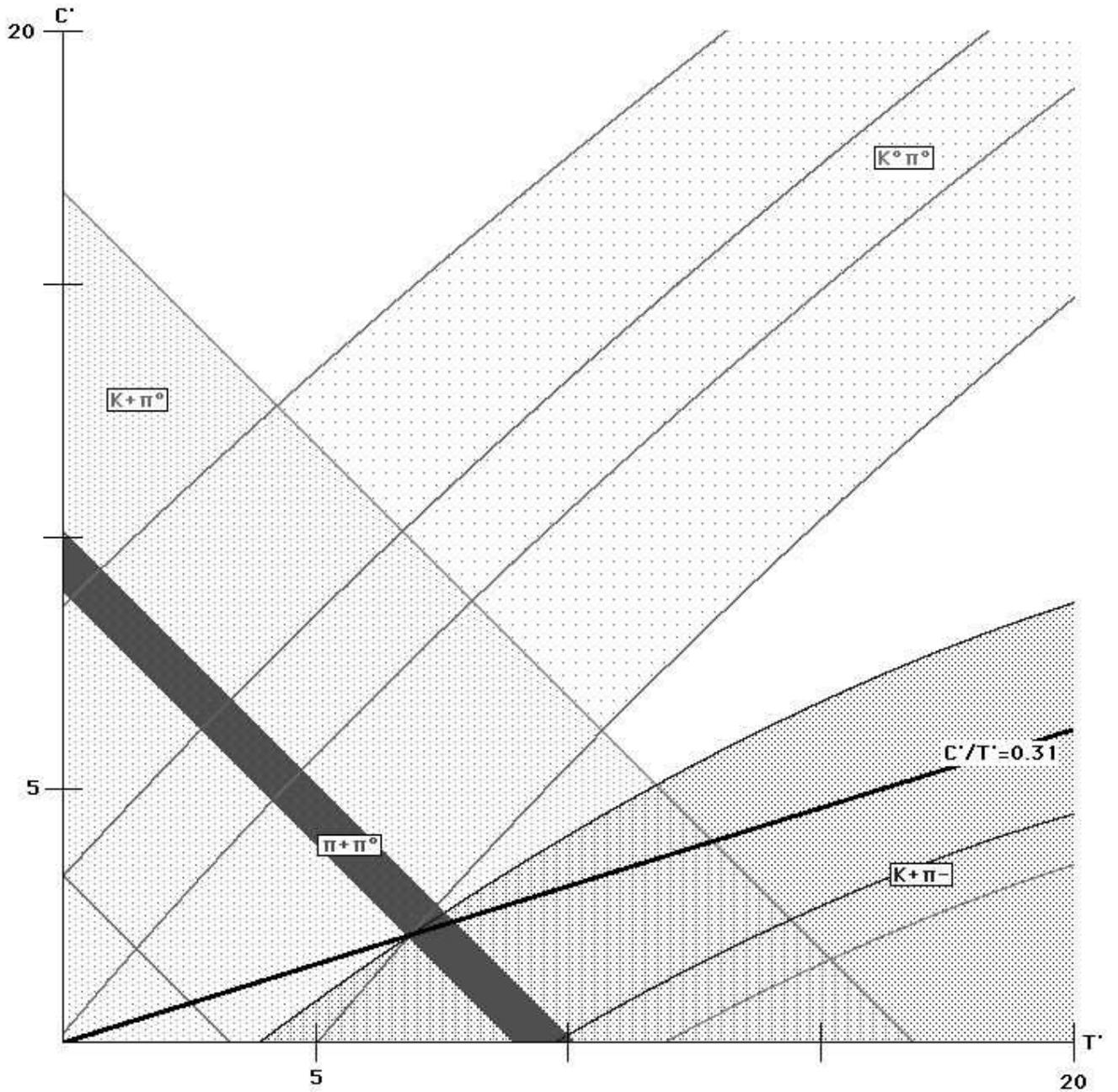}
\end{center}
\caption{\label{58pic} $[B\rightarrow K\pi]$ relations for $\gamma=58.0^\circ$}
\end{figure}

This graphical method of analysis was also applied to $B\rightarrow K\eta$ and $K\eta'$ amplitudes, graphing in terms of $S'$ and $|C'/T'|$. These decays have previously been analyzed by Chiang \emph{et al}~\cite{Chiang:2004nm}, but the recent data have not been previously analyzed.  At lower values of $\gamma$ ($38^\circ$ and above), all four amplitudes overlapped (for values of $|C'/T'|$ lower than approximately 0.4) with the exception of $[K^+\eta]$,  the amplitude of which remains too high to overlap until $\gamma$ rises to $74^\circ$, at which point all four $B\rightarrow K\eta$, $K\eta'$ amplitudes overlap (for $|C'/T'|$ lower than approximately 0.3). The overlap continues at least until $\gamma=80^\circ$. While the large values of  $\gamma$ preferred here are not inconsistent with the PDG range, they are inconsistent with the limits obtained in our earlier analysis, and suggest either that the branching ratio of $[K^+\eta]$ (and possibly $[K^0\eta]$ also) are lower than current experimental data indicate, or that the graphical decomposition is less simple than has been assumed.

Applying this analysis also to $B\rightarrow\pi\pi$ amplitudes, the current branching ratio data proves inadequate to provide any useful new constraints on $P$, although it is worth noting that consistency between the data requires a value of $|C'/T'|$ close to unity, supporting the work of Buras, Fleischer et al \cite{0312259} \cite{0402112}, and He and McKellar~\cite{He:2004ck} rather than traditional estimates.

There have been a number of analyses of the recent decay data of B mesons to charmless states~\cite{Wu:2004xx, Charng:2004ed, He:2004ck, Buras:2004th, Buras:2004dc, Baek:2004rp, Parkhomenko:2004mn}, using somewhat different methods, and different simplifying assumptions.  There is no consensus from these analyses on the size of $|C'/T'|$.  All agree with us that there are difficulties in obtaining simultaneous fits to the $K\pi$ and $\pi\pi$ decay channels, and some appeal to new physics as a way out.  None have attempted to include the $K\eta$ and $K\eta'$ decay channels in the fit, as we have done here.

In conclusion, the graphical decomposition of $B$ decay amplitudes under the framework of unbroken or partially-broken SU(3) symmetry is a useful tool, which allows insight into the fitting process. It provides a method to probe the magnitude of diagram partial amplitudes, CKM factors and complex phases, providing another check on theoretical predictions while being inclusive of multiple experimental results. As data on rare $B$ decays improves over the next decade, greater precision of results should hopefully allow this analysis to be extended to smaller, suppressed diagrams and the effects of any new physics will become easier to identify.

\begin{acknowledgments}
This work has been supported in part by the Australian Research Council.   We thank Dr A. Ignatiev and Professor X-G He for useful discussions.  BMcK wishes to thank Professor K. Terasaki for helpful discussions, and the Yukawa Institute for Theoretical Physics for their hospitality which made those discussions possible.
\end{acknowledgments}

\bibliography{bpaperb}

\end{document}